# Nanohybrids from Nanotubular J-Aggregates and Transparent Silica Nanoshells

Yan Qiao,[a,b,c]* Frank Polzer,[a,d] Holm Kirmse,[a] Stefan Kirstein,[a]* and Jürgen P. Rabe[a,b]

**Organic-inorganic nanohybrids have been synthesized by in-situ coating of self-assembled, nanotubular J-aggregates with helically wound silica ribbons. The J-aggregates retain their outstanding optical properties in the nanohybrids, but with higher mechanical stiffness, better processability, and an improved stability against photo-bleaching and chemical ambients.**

Tubular J-aggregates formed by amphiphilic cyanine dyes have attracted significant attention[1,2] since they can serve as model systems for artificial nanostructures to mimic chlorosomal light-harvesting (LH) complexes of green sulfur and non-sulfur bacteria, which are known to exhibit the highest LH efficiencies.[3] Similar to the natural LH centers, the tubular J-aggregates are comprised of dyes and display double-walled tubular structures with well-defined molecular orientations.[4] Owing to the highly ordered molecular stacking, these structures exhibit the typical optical properties of J-aggregates, such as strong, narrow, and, with respect to their monomers, red shifted absorption bands, a nearly resonant narrowband fluorescence emission, and strong exciton delocalization and migration[5,6]. However, the investigation and application of these J-aggregates has posed a long-standing challenge due to their poor stability at elevated temperatures or extreme pH values, low photostability, and weak mechanical strength. Therefore, a method that can increase the mechanical and (photo)chemical stability of the J-aggregates in solution is desirable.

Silication of functional organic building blocks is a known tool to build organic-inorganic hybrid systems and has attracted much interest, because of the optical transparency of silica in combination with biocompatibility, structural robustness, improved environmental stability, and feasibility of chemical modification.[7,8] Therefore, promising attempts were made to combine J-aggregates with silica, using silica sphere encapsulated J-aggregates or silica/J-aggregate films[9]. Mann et al. has reported template-directed synthesis of silica-coated porphyrin J-aggregate nanotapes exhibiting retention of the optical properties.[10] Within the silica host matrix, the optical properties of the J-aggregate could be modulated by tuning the coupling to localized plasmons.[11] Inspired by these previous results, we present a route to synthesize silica nanoshells onto nanotublular J-aggregates in order to enhance their photostability while preserving the optical properties. Moreover, the soft formation of a silica shell via the sol-gel method provides the possibility to transcribe the morphology of the nanotubular J-aggregates at least partially into the inorganic silica shell. Previous work has proposed a helical superstructure for the tubular J-aggregates,[12] which raises the question, whether this will be reflected by the morphology of the silica shell.

The cyanine dye 3,3′-bis(2-sulfopropyl)-5,5′,6,6′-tetrachloro-1,1′-dioctylbenzimidacarbocyanine (C8S3) is known to self-assemble into nanotubular J-aggregates in a water/methanol solution (100/13, v/v).[1,13] The resulting J-aggregates are double-walled nanotubes with an outer diameter of $13 \pm 1$ nm, an inner diameter of $6.5 \pm 1$ nm, and lengths up to tens of micrometers (Fig. S1). The UV-vis spectra show a large red-shift upon the formation of the J-aggregate. Using these supramolecular nanotubes as supporting scaffolds, optical transparent helical silica nanoshells were prepared through the sol-gel process with 3-aminopropyltriethoxysilane (APTES), and tetraethoxysilane (TEOS) as silica precursors.

Cryogenic transmission electron microscopy (cryo-TEM) was employed to visualize the silica-coated J-aggregates in-situ.[12] As shown in Fig. 1a, thread-like structures were obtained, which are deemed as silica nanoshell coated J-aggregate hybrids. Cross-sectional line profiles were taken across the nanohybrids as presented in Fig. 1b. An overlay of line scans taken from different hybrid nanotubes resolves the average diameter of the nanohybrids to be $21 \pm 1.5$ nm, as measured by the full width at half maximum (FWHM).

The line scans show similar diameters manifesting the homogeneity of the hybrid nanotubes. Compared to the diameter of the bare J-aggregate of 13 ± 1 nm, the mean diameter of silica coated J-aggregates has increased by 7 - 8 nm. This increase is assigned to a silica shell with a mean thickness of 3.5 to 4 nm. The growth of the silica shell on the outer surface of tubular J-aggregates is illustrated further by Fig. 1c. An end section of the nanohybrid is shown in a magnified view. The remaining of the original tubular J-aggregate can be seen as it protrudes from the silica shell. The diameters of the J-aggregate tube are the same as observed for the bare J-aggregates (13 nm outer diameter and 6.5 nm inner diameter).[2] Obviously, the sol-gel process does not cause significant changes in morphology or size of the J-aggregate nanotubes. This is further supported by optical spectroscopy, as shown below. In Fig. 1d, a top view of the open end of an upstanding nanohybrid is captured accidentally. The center of the tube is filled with a high contrast material that may be identified as filling of the tube with silica. The diameter is close to what was previously measured in the J-aggregates filled with silver.[14]

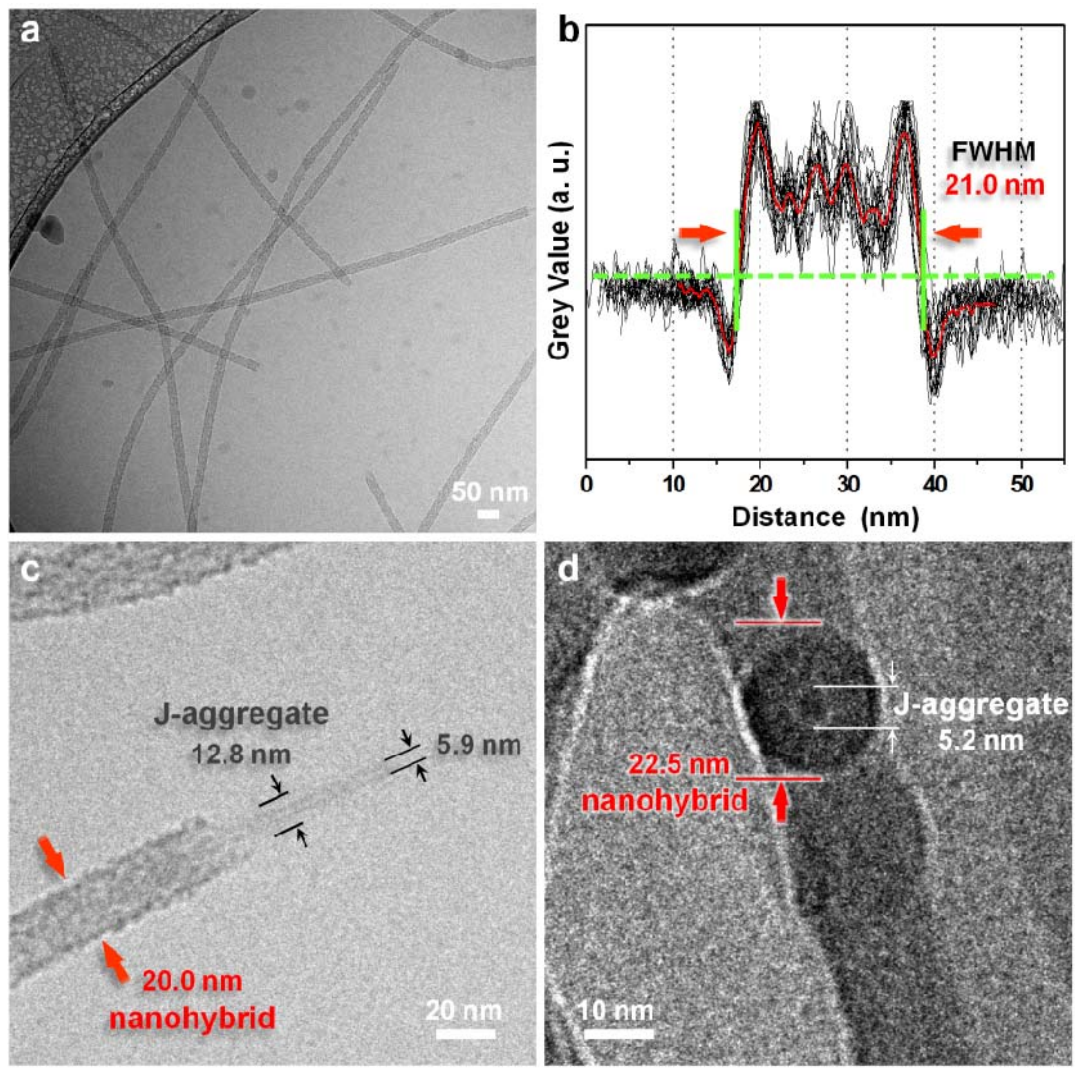

*Fig. 1 Cryo-TEM images and analysis of silica helical nanoshell coated J-aggregates. (a) An overview of nanohybrids. (b) Overlaid line scans across different individual nanohybrids reflecting the highly structural homogeneity. The red curve is the average curve which displayed the average diameter was 21 ± 1.5 nm. The four peaks in the red curve proved the repeated periods of the helical silica nanoshells. (c) An end section of the nanohybrid denoting the remaining of the original tubular J-aggregate. (d) An upstanding nanohybrid indicating the inner tube was filled by silica.*

As verified in Fig. S2, the nanohybrids with a high product yield can extend in length up to tens of micrometers. Negligible amounts of non-templated silica were noted. It is important to note that no obvious bundling effect of the J-aggregate is observed after the silication with APTES/TEOS and most of the nanohybrids are isolated. Additionally, the deposition of the silica nanoshells increased the stiffness of the nanotubes, as is concluded from the typical increase of bending radius. High resolution-TEM (HR-TEM, Fig. S3) and energy dispersive X-ray spectroscopy (EDXS, Fig. S4) measurements support that the nanostructures are comprised of J-aggregates (origin of S-signal) and silica (origin of Si-signal).

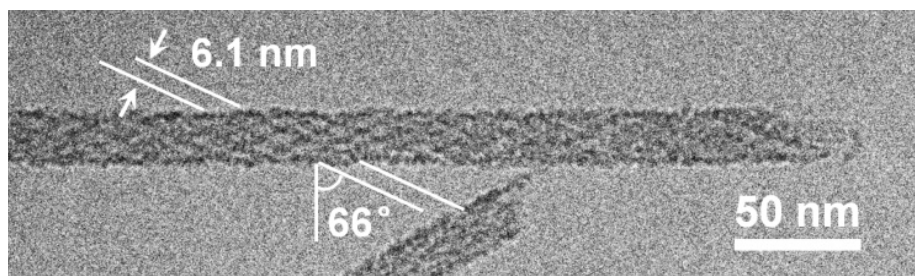

*Fig. 2 Magnified view of a cryo-TEM image of single silica nanoshell coated J-aggregates. The width and the helix angle of the silica ribbons are indicated.*

In a magnified cryo-TEM image (Fig. 2), a typical silica/J-aggregate nanohybrid revealed a well-ordered net-like pattern which is indicative of a repetitive pattern expected from the projection image of a helical structure. A close inspection reveals the width of the ribbons that constitute the helix is 6.1 ± 0.3 nm with a helix angle of 66 ± 2 °. The helical handedness, however, cannot be identified from the projection images obtained by cryo-TEM. A sketch of a proposed 3D model of the nanohybrids using the parameters extracted from Fig. 2 is shown in the ESI (Fig. S5). It is important to note that the width of these ribbons as well as the tilt angle differ from the values proposed for the pure J-aggregates[12], which means the silica super structure is not a direct one-to-one transcript of the aggregate structure. The formation of the silica nanoshell is initiated and driven by electrostatic forces that require charge matching between silica precursor and template (Fig. S6). [15,16] The surface of the tubular J-aggregates is assumed to be negatively charged due to the sulfonate groups of the C8S3 anions. The cationic species, such as [Si(OH)3(NH3)]+ hydrolyzed from ATPES, are attracted by the negative surface potential, forming a ultrathin silica layer.

Further condensation of hydrolysis products of TEOS, e. g. [Si(OH)4], leads to the formation of the silica nanoshell specifically at the J-aggregate surface that was previously covered by APTES. It is supposed that the silica nanoshell can inherit and amplify small corrugations of the aggregate surface. Therefore, the helical winding of the ribbons is proposed to be induced by the helical superstructure of the tubular aggregates.[12] The optical properties of the silica coated J-aggregates were investigated by UV-vis and fluorescence measurements (Fig. 4a), which show that encapsulation of the J-aggregates by silica does not lead to significant changes of the spectrum and hence reorganization of the J-aggregate structure. The very minor differences may be either due to the different polarizability[17] of the close environment of the dyes, or due to very slight structural change. Since in cryo-TEM images all J-aggregates are covered by silica, the spectra truly represent the spectra of silica covered J-aggregates. Fig. 3b illustrates uncorrected fluorescence spectra showing that the emission of silica nanoshell-coated J-aggregates remains unchanged besides some difference in intensity. Using rhodamine B as a standard, the fluorescence quantum yield ($\Phi_f$) of silica nanoshell-coated J-aggregate ($\Phi_f$ = 8%) was determined to be approximately 13% higher than that of bare J-aggregates ($\Phi_f$ = 6%) (Fig. S7).

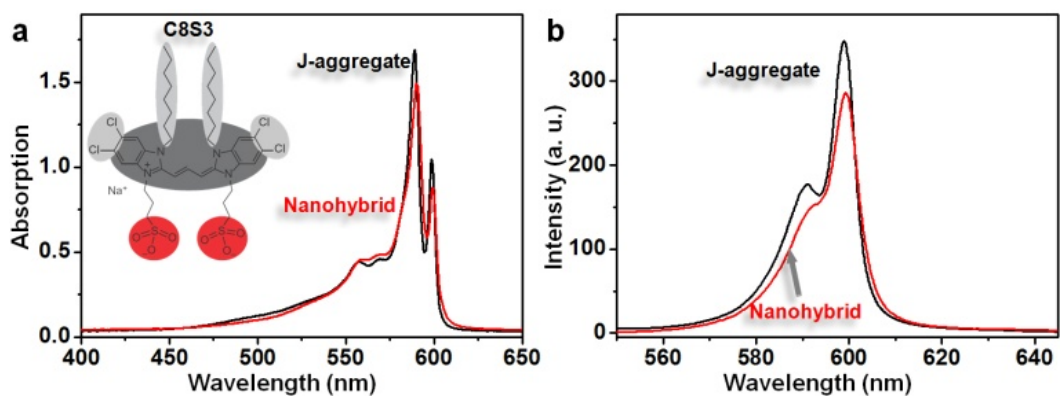

*Fig. 3 Optical properties of nanohybrids and J-aggregates of the dye C8S3. (a) Absorption spectra; (b) PL spectra of as-prepared nanohybrid and J-aggregate ($\lambda_{ex}$= 500 nm)*

The thickness of the silica nanoshell can be facilely adjusted by varying the concentration of the silica precursor. The thickness of the silica nanoshell on the J-aggregate was found to increase linearly with TEOS in the concentration range of 3.6 mM to 14.3 mM with an increase per mM of 0.5 nm (Fig. S8, S9). It is therefore proposed that this method is a straightforward, effective, and adaptive way to synthesize silica nanoshell coated J-aggregate with defined shell thickness.

Preliminary experiments indicate that the silica nanoshell stabilizes the J-aggregates against photobleaching. The nanohybrid and J-aggregate samples were irradiated continuously with 500 nm light and the fluorescence intensities at 600 nm were recorded vs. time. As shown in Fig. S10, after 60 min the fluorescence emission of the nanohybrids was bleached to approx. 60%, while, under same conditions, the fluorescence of uncovered J-aggregates was bleached to almost 10% of the original intensity. Since photobleaching of cyanine dyes is known to be caused by a chemical reaction with singlet oxygen or hydroxyl radicals, we conclude that the silica nanoshell is dense enough to significantly reduce the penetration of such radicals. However, exact quantification of photobleaching turned out to be difficult due to variations between individual samples (Fig. S8) and is therefore beyond this communication.

Conclusions

An in-situ synthesis of silica nanoshells on nanotubular J-aggregates was developed based on the electrostatic attraction between silica precursors and the J-aggregates. The resulting nanohybrids exhibit a helical superstructure, induced by the supramolecular helical structure of the J-aggregates. The nanometer sized silica shell provides higher mechanical stiffness, better processability, and an improved stability against chemical ambients and photobleaching. All these factors make this hybrid nanomaterial very promising for advanced optical spectroscopy studies, and also for the hierarchical build-up of robust and complex functional systems for light harvesting.

## Notes and references

a Department of Physics, Humboldt-Universität zu Berlin, Newtonstr 15, 12489 Berlin, Germany.

b IRIS Adlershof, Humboldt-Universität zu Berlin, Zum Großen Windkanal 6, 12489 Berlin, Germany.

c present address: Centre for Protolife Research and Centre for Organized Matter Chemistry, School of Chemistry, University of Bristol, Bristol BS8 1TS, United Kingdom.

d present address: Materials Science & Engineering, University of Delaware, Newark, DE 19716, USA.

† Acknowledgements: We thank Dr. Omar Al-Khatib and Evi Poblenz for fruitful discussions and assistance, respectively, and we acknowledge funding of the DFG through CRC 951. Dr. F. Polzer acknowledges funding also through the Joint Lab for Structural Research Berlin within IRIS Adlershof.

Electronic Supplementary Information (ESI) available: materials and methods, TEM and HRTEM images of nanohybrids, EDXS analysis, measurement of fluorescence quantum yields, and photo bleaching experiments are presented. See DOI: 10.1039/c000000x/

# Supporting Information

## 1. Materials and methods

**Preparation of tubular J-aggregates.** Double-walled J-aggregate nanotubes were prepared in water/methanol (9:1, v/v). Stock solution of C8S3 (2.92 mM) was prepared by directly dissolving C8S3 powder (FEW Chemicals, Germany) in methanol (≥99.9% GC, Sigma-Aldrich). Then 130 μL of the C8S3 stock solution was added to 500μL of ultrapure $H_2O$ followed by thorough vortex to ensure complete mixing. An immediate color change of the solution (from orange to deep pink) was observed, indicating the formation of tubular J-aggregates. The solution was stored in the dark for 24 h before adding an additional 500 μL of ultrapure $H_2O$, which gave a final concentration of 0.336 mM. Fresh J-aggregate solutions were stored in the dark and used within three days after preparation.

**Preparation of silica nanoshell on the C8S3 J-aggregate.** For the helical silica shell synthesis, typically, 2 μL of 3-aminopropyltriethoxysilane (APTES, sigma, 99%) methanol (v/v = 1:49) solution and 15 μL tetraethoxysilane (TEOS, sigma, 99%) methanol (v/v = 1:49) solution was added to a 250 μL C8S3 J-aggregate solution. The solution was vortexed to ensure complete mixting and stored for 24 hours before measurement. All the experiments were finished under red light and the samples were stored in dark to protect from photobleaching.

**Absorption spectroscopy.** Absorption spectra of the solution were taken with a double-beam UV-vis spectrometer (Shimadzu UV-2101PC, Duisberg, Germany) in a 0.2 mm, 1.0 mm, or 10.0 mm demountable quartz cell (Hellma GmbH, Germany).

**Fluorescence spectroscopy**. Fluorescence spectra were taken on a fluorescence spectrometer (JASCO FP 6500) in quartz cells (Hellma GmbH, Germany) with the path length of 0.2 mm, 1.0 mm, or 10.0 mm. The fluorometer came with built-in light-source (Xe-lamp UXL-159, 150W), two grating-monochromators with variable slit-width and detector. The photobleaching experiments were conducted within this machine at a wavelength of 500 nm (slit width = 10 nm).

**Cryogenic transmission electron microscopy (Cryo-TEM).** Droplets of the solution (5 μL) were applied to carbon film (1 μm hole diameter) covered 200 mesh grids (R1/4 batch of Quantifoil Micro Tools GmbH, Jena, Germany), which had been hydrophilized before use by plasma process. Supernatant fluid was removed by Vitrobot Mark (FEI company, Eindhoven, Netherlands) until an ultrathin layer of the sample solution was obtained spanning the holes of the carbon film. The samples were immediately vitrified by propelling the grids into liquid ethane at its freezing point (90 K). Frozen samples were transferred into a JEOL JEM2100 (JEOL, Eching, Germany) using a Gatan 914 high-tilt cryo-transfer holder and Gatan work station (Gatan, Munuch, Germany). Microscopy was carried out at a 94 K sample temperature using the microscope's low-dose protocol at an accelerating voltage of 200 kV (LaB6-illumination). The defocus was chosen to be approx 2 μm to create sufficient phase contrast for imaging. Image analysis was performed with Image J software[1].

**High Resolution Transmission Electron Microscopy (HR-TEM).** High resolution transmission electron microscopy (TEM) was performed with a JEOL 2010F operating at 200 kV. Droplets of the sample (5 μL) were applied to carbon film coated 200 mesh grids (Quantifoil Micro Tools GmbH, Jena, Germany), which had been hydrophilized before use by simply storing the grids over a water bath for a day. The excess fluid was removed with a filter paper to form an ultra-thin layer on the carbon film. Image analysis was performed with Image J software[1].

## 2. Sketch and spectra of C8S3 J-aggregates

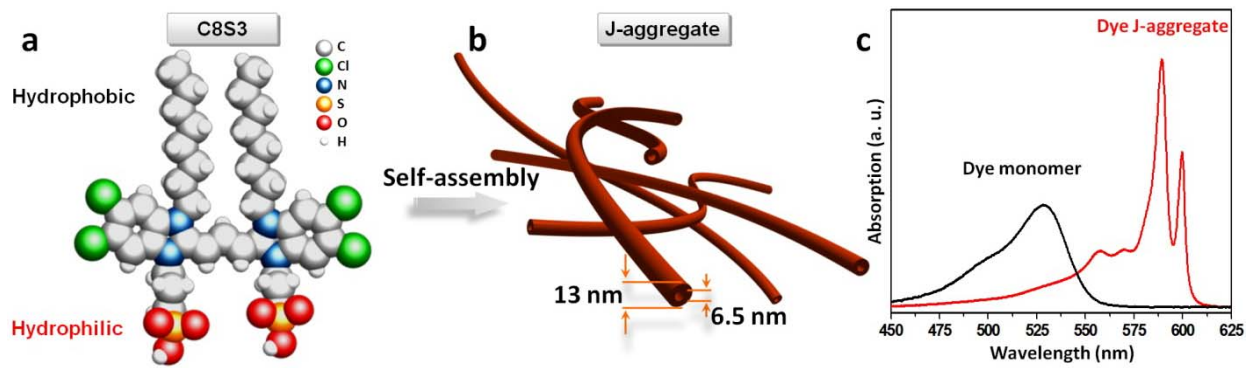

*Figure S1.* Sketch of molecular self-assembly into J-aggregates. (a) Molecular structure of amphiphilic cyanine dye C8S3; (b) sketch of nanotublular J-aggregates; (c) absorption spectra of C8S3 monomer (black) and J-aggregate (red) solution, in which J-aggregates exhibit narrow and red-shifted absorption peaks relative to monomers.

## 3. TEM image of silica nanoshell coated J-aggregate hybrids

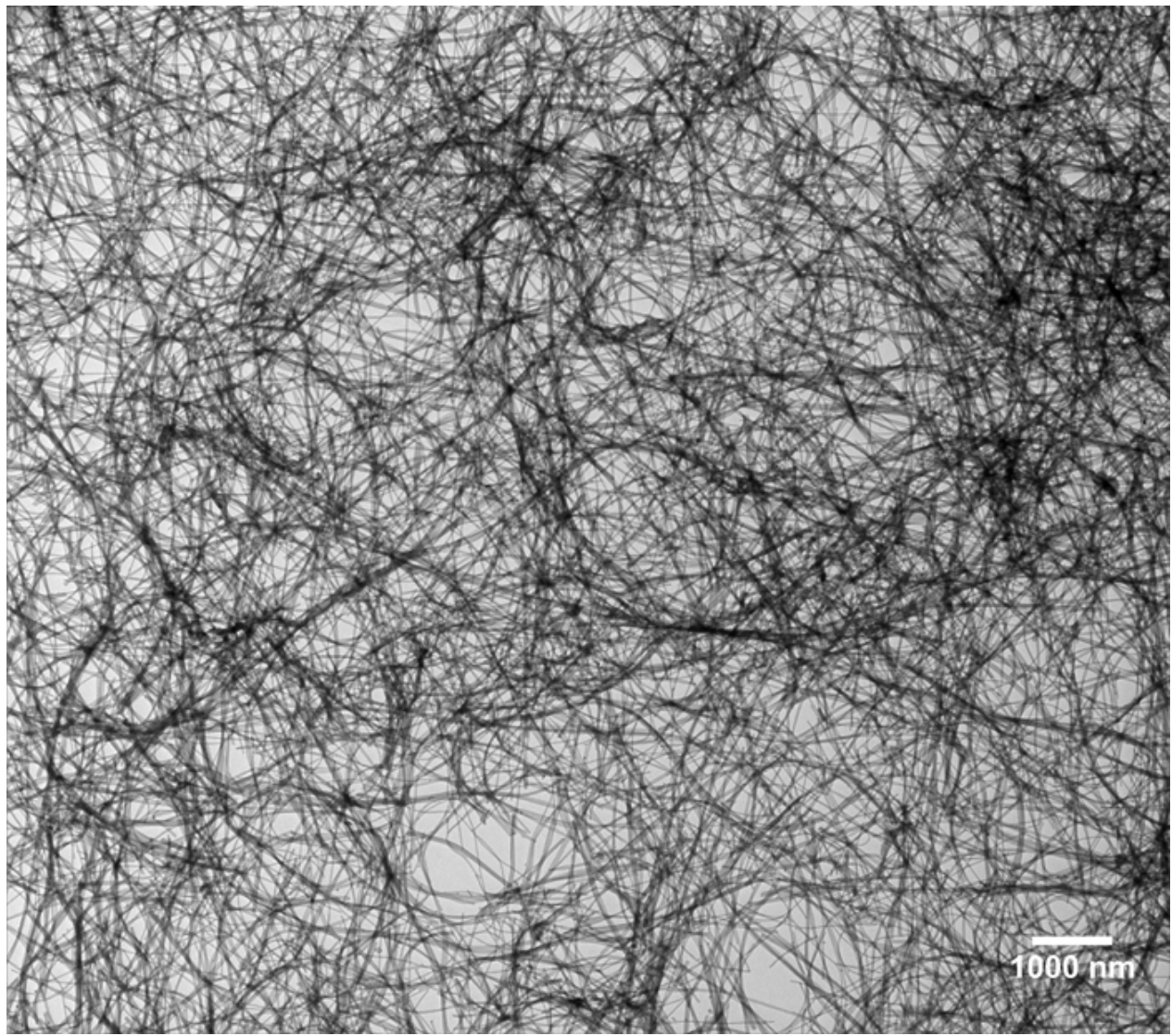

*Figure S2.* TEM image of silica nanoshell coated J-aggregate nanohybrids. The image shows tubular nanohybrids are several micrometers long.

**4. HRTEM image of silica nanoshell coated J-aggregate hybrids**

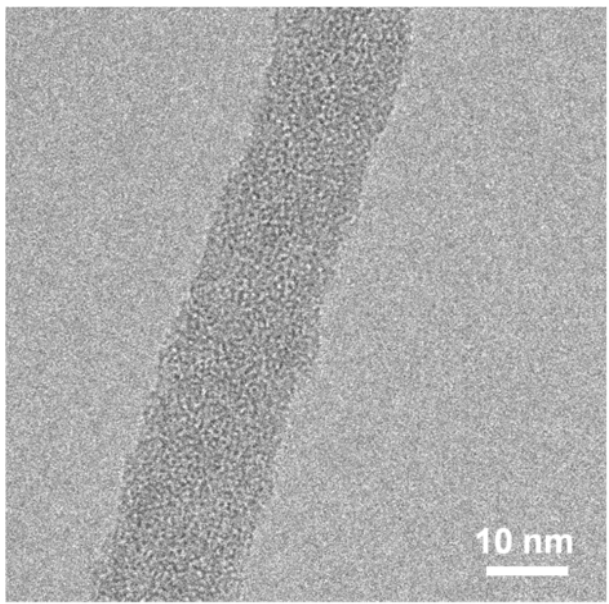

*Figure S3.* HR-TEM image of the as-prepared silica nanoshell-coated J-aggregate showing porous amorphous structure.

## 5. EDXS analysis of silica nanoshell coated J-aggregate hybrids

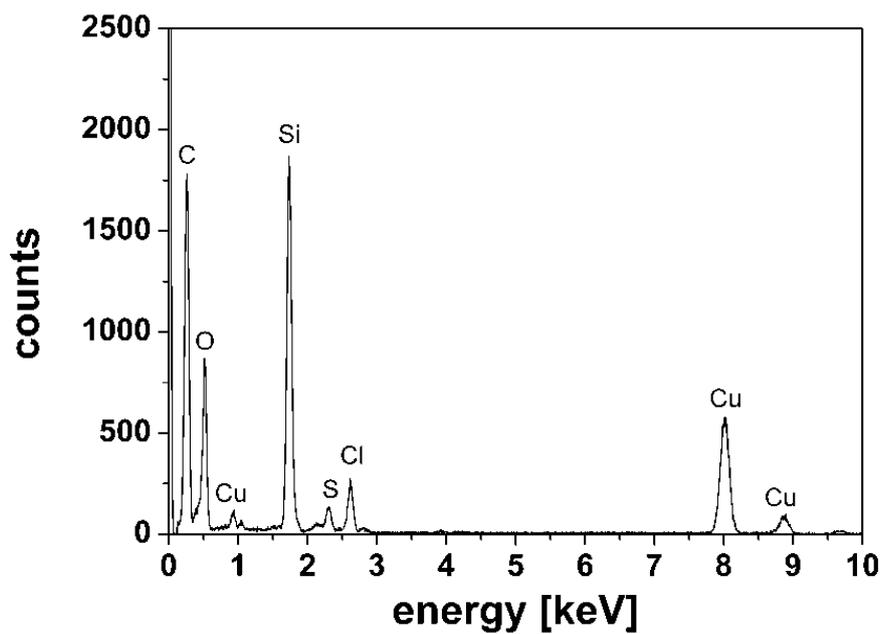

*Figure S4.* EDXS analysis of the as-prepared silica nanoshell-coated J-aggregate nanohybrids confirms the presence of Si and S elements.

**6. Sketch of electrostatically driven silication of tubular J-aggregate**

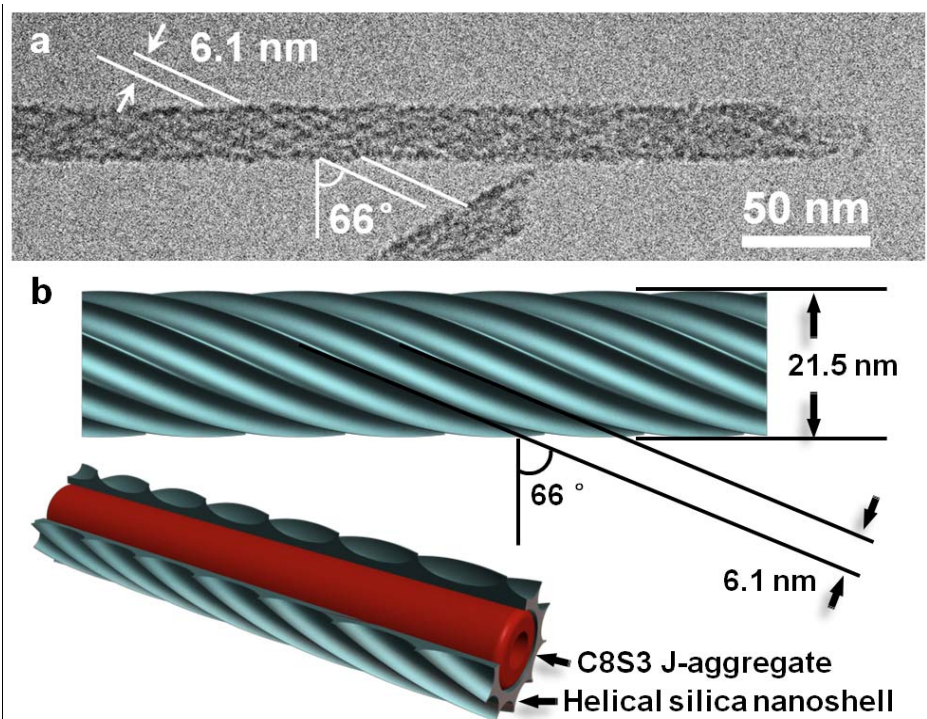

*Figure S5.* (a) Cryo-TEM image of silica nanoshell coated J-aggregates with detailed analysis of the superstructure. (b) 3D model of the surface and cross-section of silica coated J-aggregates exhibiting the parameters of helical structures.

**7. Sketch of electrostatically driven silication of tubular J-aggregate**

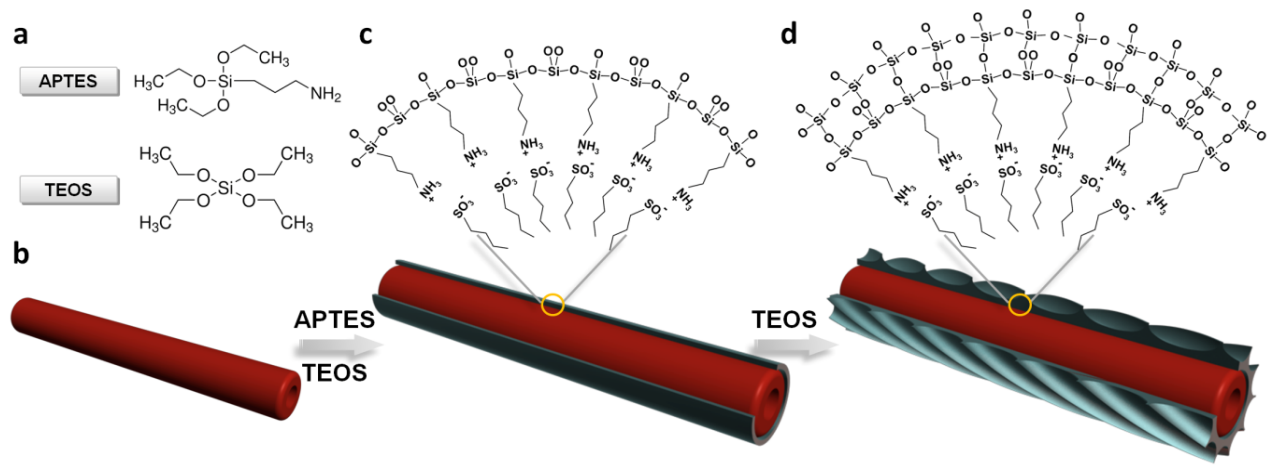

*Figure S6.* Sketch of electrostatically driven silication of tubular J-aggregate. (a) Molecular structures of the constituents: APTES, and TEOS; (b) sketch of J-aggregate nanotube, (c) electrostatic adsorption of APTES on J-aggregates; (d) condensation of TEOS on J-aggregate nanotubes.

## 8. Measurement of fluorescence quantum yields of J-aggregate with and without silica shell

The fluorescence quantum yield ($Q$) is obtained using the comparative method of Williams *et al.*, which involves the use of well characterized standard fluorophore with known $Q$ values. The quantum yield of the test sample is calculated using:

$$Q_X = Q_{ST}\left(\frac{I_X}{I_{ST}}\right)\left(\frac{OD_{ST}}{OD_X}\right)\left(\frac{\eta_X^2}{\eta_{ST}^2}\right)$$

Where the subscripts ST and X denote standard and test respectively, $Q$ is the fluorescence quantum yield, $I$ is the integrated fluorescence intensity, $\eta$ is the refractive index, and $OD$ is the optical density. In this work, a series of solutions were prepared with optical densities between 0.1 and 0.01, and a more accurate method was used to determine $Q$ by the following formula:

$$Q_X = Q_{ST}\left(\frac{Grad_X}{Grad_{ST}}\right)\left(\frac{\eta_X^2}{\eta_{ST}^2}\right)$$

where *Grad* is the gradient obtained from the plot of integrated fluorescence intensity $I$ versus the absorbance. Fluorescein in 0.1 M NaOH solution ($Q$ = 95%)[2] Rhodamine B ($Q$ = 31%)[3] was used as a standard sample for J-aggregate. All the fluorescence spectra were recorded with constant slit widths. The obtained quantum yields of J-aggregate with and without silica shell were 8%, and 6%, respectively.

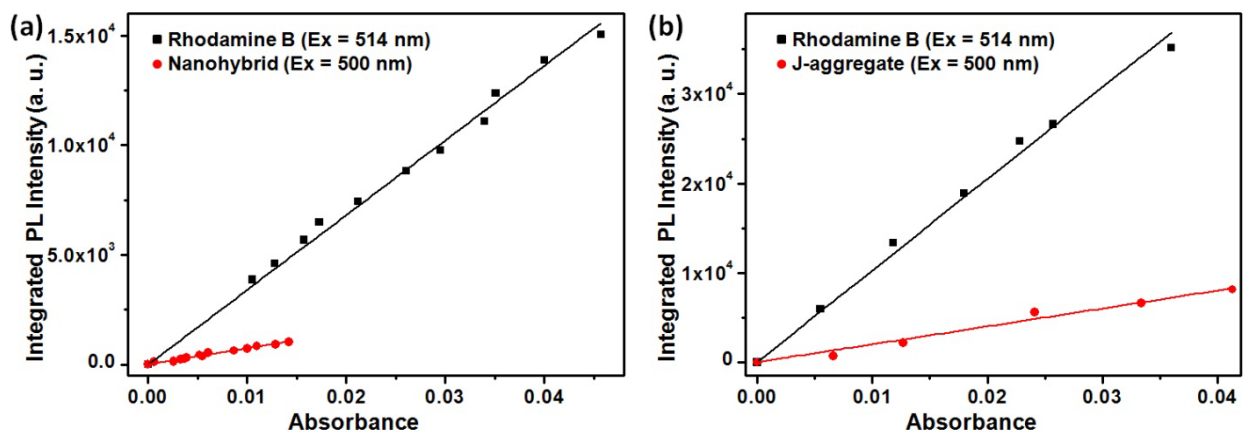

*Figure S7.* Linear plots of the measured absorption and integrated fluorescence intensity of: (a) rhodamine B and C8S3 J-aggregate; (b) rhodamine B and silica nanoshell-coated C8S3 J-aggregate.

## 9. Adjustable thickness of the silica nanoshell by varying the concentration of the silica precursor

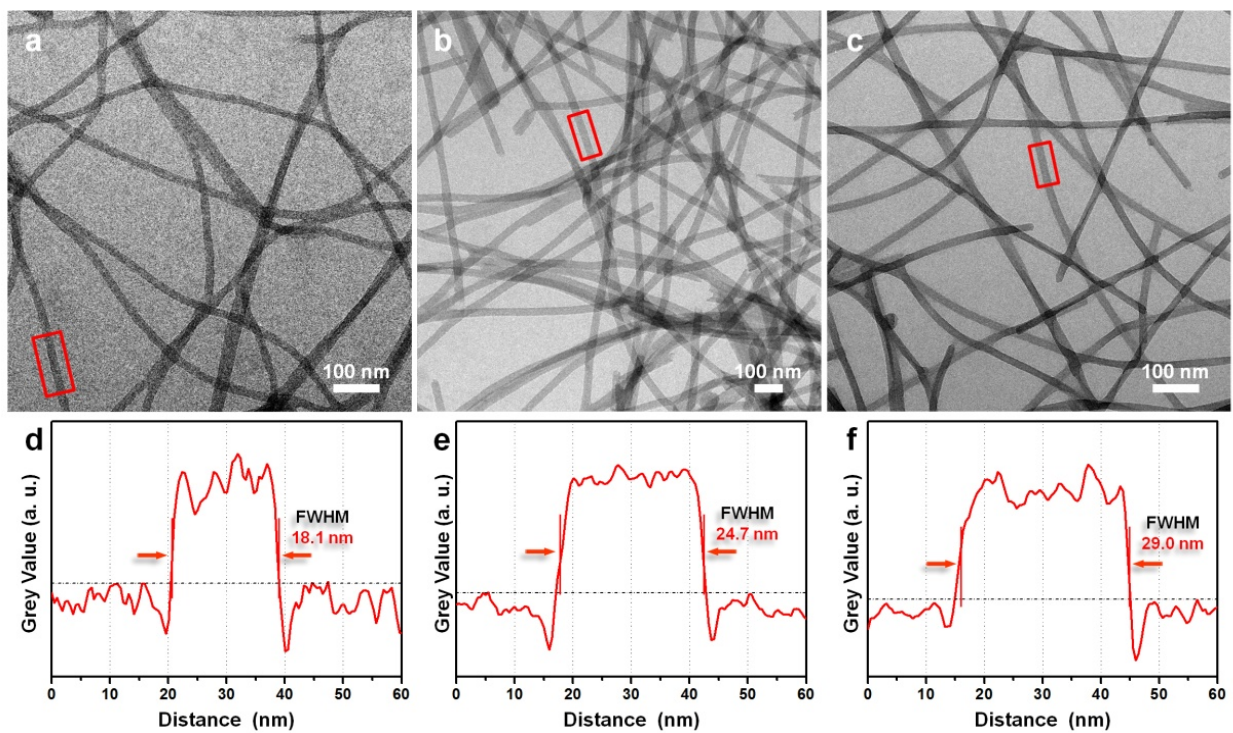

*Figure S8.* TEM images (a-c) and line scans (d-f) of silica coated nanohybrids prepared with different ATPES/TEOS concentration. (a, d) 0.684 mM / 3.58 mM; (b, e) 0.684 mM / 10.7 mM and (c, f) 0.684 mM / 14.3 mM.

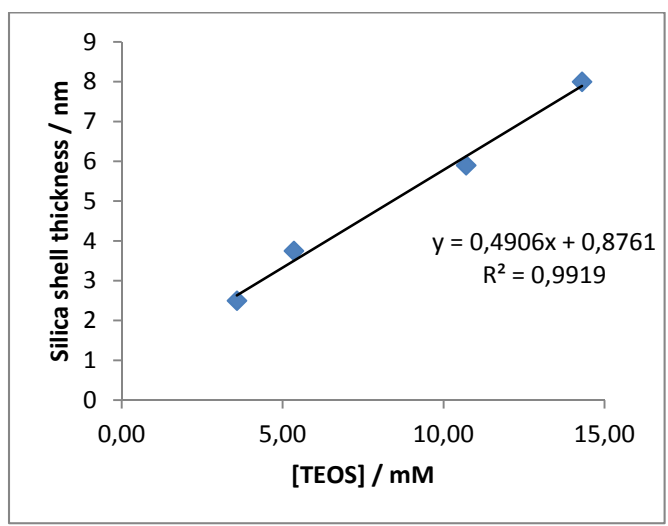

*Figure S9.* Plot of silica shell thickness versus concentration of TEOS. The shell thickness was calculated from the FWHM values of the total diameter of the J-aggregate/silica hybrid (as shown in Fig S6), by subtracting 13 nm for the pure J-aggregates, and division by two.

## 10. Preliminary photo-bleaching experiments

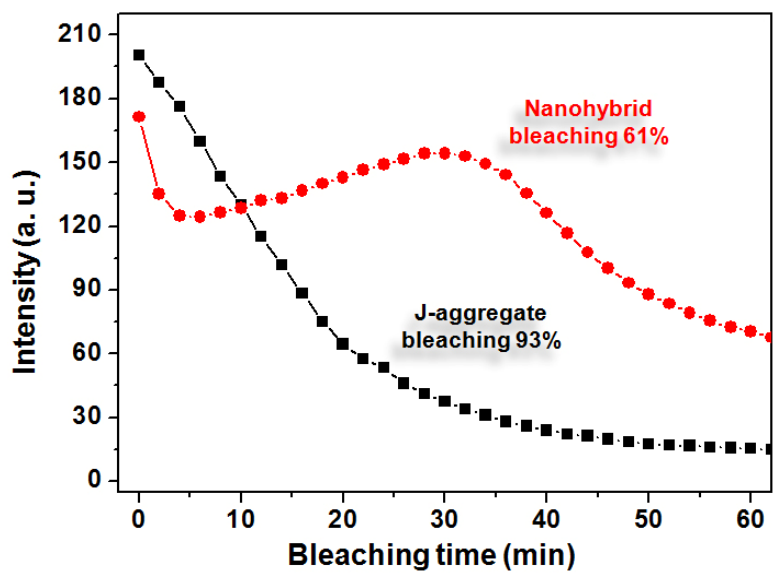

*Figure S10.* PL intensities of nanohybrids and J-aggregate at 600 nm versus photo bleaching time recorded under radiation with Xe lamp (from fluorescence spectrometer), spectra taken at time intervals of 2 min ($\lambda_{ex}$= 500 nm).